# Watt parameters for the Los Alamos Model : Subroutine *getab*

J. P. Lestone
X-1-TA, Los Alamos National Laboratory
September 5$^{th}$, 2007

**Abstract**

Many neutron transport Monte-Carlo codes can randomly sample fission neutron energies from a Watt spectrum. The quality of simulations depends on how well the Watt spectrum represents the true energy spectrum of the fission neutrons, and on one's choice of the Watt parameters *a* and *b*. The energy spectra of fission neutrons have been calculated and tabulated for the neutron induced fission of $^{235,238}$U and $^{239}$Pu as a function of incoming neutron energy by Madland using the Los Alamos Model. Each of these energy spectra are mapped into time-of-flight space and fitted with a Watt spectrum. A subroutine *getab* has been written to interpolate these results, so that Watt *a* and *b* parameters can be estimated for all incoming neutron energies up to ~16 MeV.

**Introduction**

Neutron transport in the presence of fission requires knowledge of the energy spectra of fission neutrons. Many Monte-Carlo codes can randomly sample fission neutron energies, $\varepsilon$, from a Watt spectrum

$$f(\varepsilon) = \frac{2\exp(-ab/4)}{\sqrt{\pi a^3 b}} \exp(-\varepsilon/a) \sinh(\sqrt{b\varepsilon}) \quad . \tag{1}$$

Different sampling techniques can be employed. The method used by MCNP [1] is (in Fortran)

```
        x=1.+0.125*a*b
        wc(1)=x+sqrt(x**2-1.)-1.
        wc(2)=a*b*(wc(1)+1.)
        wc(3)=a*(wc(1)+1.)
20      t1=-log(ranf(0.d0))
        if ((log(ranf(0.d0))+wc(1)*(1.+t1))**2.gt.wc(2)*t1) goto 20
        erg=wc(3)*t1
```

where ranf(0.d0) is a random number from 0 to 1, and erg is the sampled neutron energy. The quality of simulations depends on how well the Watt spectrum represents the true energy spectrum of the fission neutrons and on one's choice of the parameters *a* and *b*. The energy spectra of fission neutrons have been calculated and tabulated for the neutron induced fission of $^{235,238}$U and $^{239}$Pu as a function of incoming neutron energy by Madland [2] using the Los Alamos Model [3]. Monte Carlo codes could sample directly from the energy spectra tabulated in ref. [2]. However, this would require that the results tabulated in ref. [2] be made available to Monte Carlo codes and the writing of a significant piece of new software. Here, we choose to take advantage of Watt spectrum sampling features available in many neutron transport codes and fit each of the energy spectra tabulated in ref. [2] with a Watt spectrum in time-of-flight space. A subroutine *getab* has been written to interpolate these results, so that Watt *a* and *b* parameters can be estimated for all incoming neutron energies up to ~16 MeV.

**Los Alamos Model fission neutrons**

Los Alamos Model (LAM) [3] calculations for the energy spectra of fission neutrons have been performed and tabulated for neutron induced fission of $^{235}$U, $^{238}$U and $^{239}$Pu for a number of inducing neutron energies from zero to ~16 MeV [2]. Fig. 1 shows Los Alamos Model calculations [2] for the energy spectra of fission neutrons from thermal neutron induced fission of $^{235}$U, and $^{239}$Pu, and 2 MeV and 14 MeV neutron induced fission of $^{239}$Pu. These spectra all peak at ~1 MeV. The main difference is in the high energy tail beyond 4 MeV. The differences in the high energy tail are easily understood within the framework of an evaporative model. Fission reactions with a higher neutron multiplicity have fission fragments at higher excitation energy and thus a higher nuclear temperature. This higher temperature hardens the energy spectrum of the evaporated neutrons.

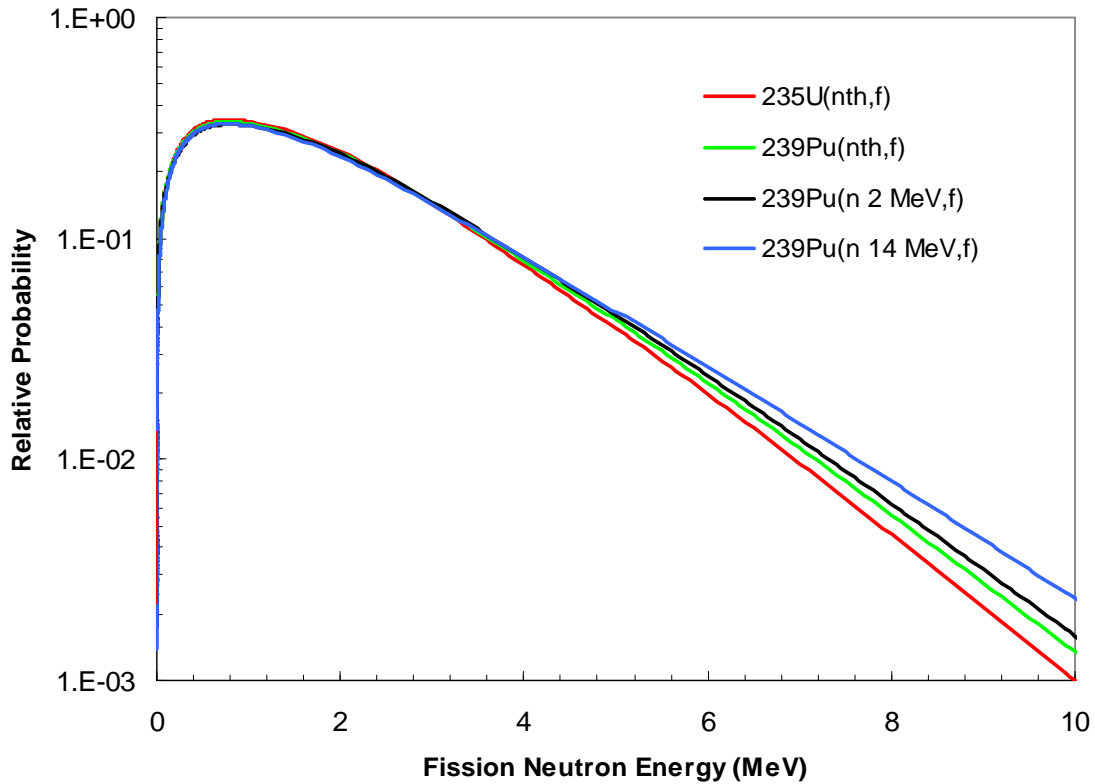

Fig. 1. Los Alamos Model calculations [2] for the energy spectra of fission neutrons for selected reactions.

For many experiments involving the neutron emission from the prompt fission of actinides, the relevant measured quantity is the neutron time-of-flight. It is therefore, instructive to plot the neutron time-of-flight spectra. Figs 2 and 3 show Los Alamos Model calculations [2] for time-of-flight spectra of fission neutrons from thermal neutron

xxLA-UR-07-6090

induced fission of $^{235}$U, and $^{239}$Pu, and 2 MeV and 14 MeV neutron induced fission of $^{239}$Pu. The assumed flight path is ~600 cm. These time-of-flight spectra are very similar at late times, and all peak near 250 ns. However, the height and precise location of the peak and the location of the leading edge are affected by the hardness of the energy spectrum. A change from thermal neutron induced fission of uranium to fast neutron induced fission of plutonium corresponds to a 5-10 ns shift in the leading edge of the time-of-flight spectrum (assuming a 600 cm flight path).

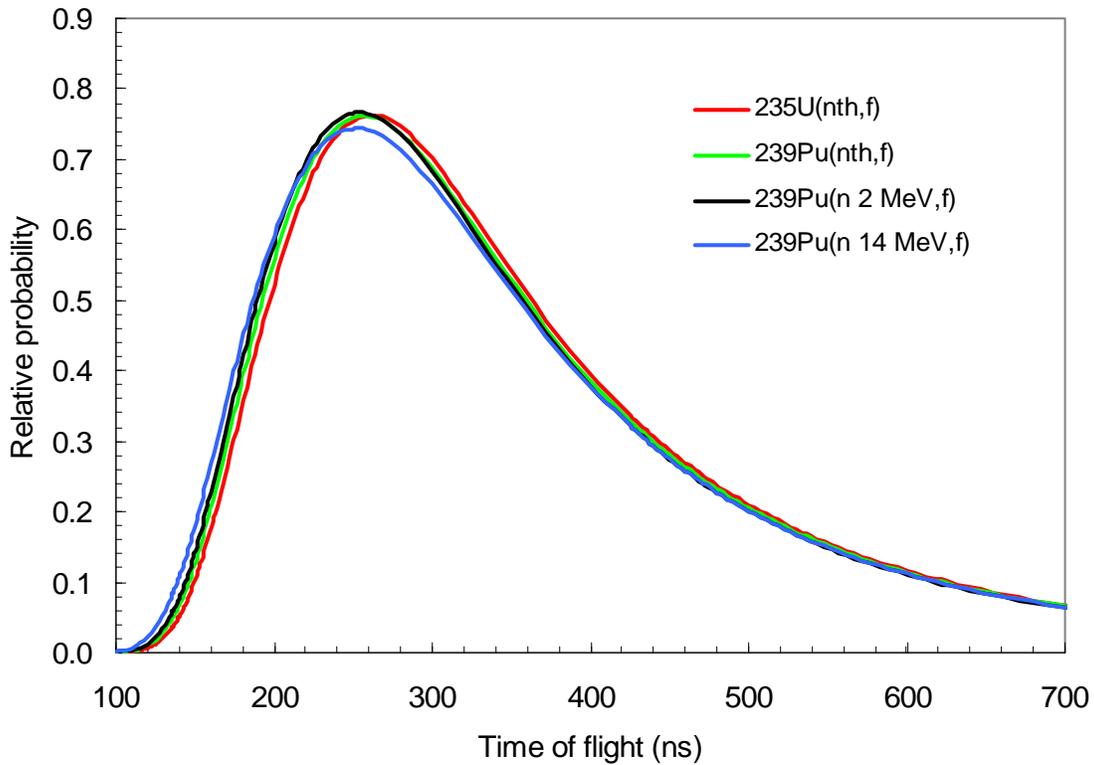

Fig. 2. Los Alamos Model calculations [2] for time-of-flight spectra of fission neutrons for selected reactions. The assumed flight path is ~600 cm.

**Watt fits**

As discussed above, for many experiments involving the neutron emission from the prompt fission of actinides, the relevant measured quantity is the neutron time-of-flight. We map the neutron energy into time-of-flight using the expression

$$t = \frac{600(\mathrm{ns}\cdot\mathrm{MeV}^{1/2})}{\sqrt{2\varepsilon}}. \qquad (2)$$

This assumes an ~600 cm flight path and non relativistic neutrons. For neutrons with energies less than 14 MeV the relativistic corrections are small. Given Eq. (2) the probability per unit energy is mapped into time-of-flight space using the expression

$$F(t)dt = f(\varepsilon)\varepsilon^{3/2}dt. \qquad (3)$$

For each of the tabulated spectra [2], Watt fitting parameters $a$ and $b$ were obtained by minimizing the chi-squared

x3

LA-UR-07-6090

$$\chi^2 = \sum_i (F_{\text{LAM}}(t_i) - F_{\text{Watt}}(t_i,a,b))^2 \ . \tag{4}$$

The summation was over time-of-flight values from 113 ns (14 MeV) to 700 ns (0.37 MeV). The fitting parameter $b$ was constrained to be larger than 0.01 MeV$^{-1}$. Fig. 4 compares the Los Alamos Model 2-MeV neutron induced fission of $^{239}$Pu time-of-flight spectrum to the corresponding Watt fit. Fig. 5 shows the corresponding ratio of the Watt fit to the Los Alamos Model. For time-of-flight values later than ~150 ns ($\varepsilon < 8$ MeV) the maximum difference between the LAM and the Watt fit is 3%. Equally good quality fits were obtained at all inducing neutron energies. The discrepancy between the LAM and the Watt fit increases to ~5% at $t$~140 ns ($\varepsilon \sim 9$ MeV) and ~10% at $t$~130 ns ($\varepsilon \sim 11$ MeV). For many applications where the signal from fission neutrons with energies greater than 9 MeV is small, the Watt fits presented here are likely to be adequate. Our Watt fitting parameters are listed in Table I and shown in Figs 6 and 7. A subroutine *getab(ia,en,a,b,wc)* (see appendix A) has been written to interpolate these results, so that Watt $a$ and $b$ parameters can be estimated for all incoming neutron energies up to ~16 MeV. The input variables are the mass number of fissioning isotopes *ia* and the energy of the inducing neutron *en* in MeV. For the reactions $^{235}$U(n,f), $^{238}$U(n,f), and $^{239}$Pu(n,f), *ia* should be set to 235, 238, and 239, respectively. If *ia* is not either 235 or 238, the result defaults to the $^{239}$Pu(n,f) reaction. The output variables are $a$, $b$, and *wc*(1:3). The *wc*(1:3) are the Watt coefficients used in MCNP [1] to sample the Watt distributions (see the introduction) and are simple functions of $a$ and $b$.

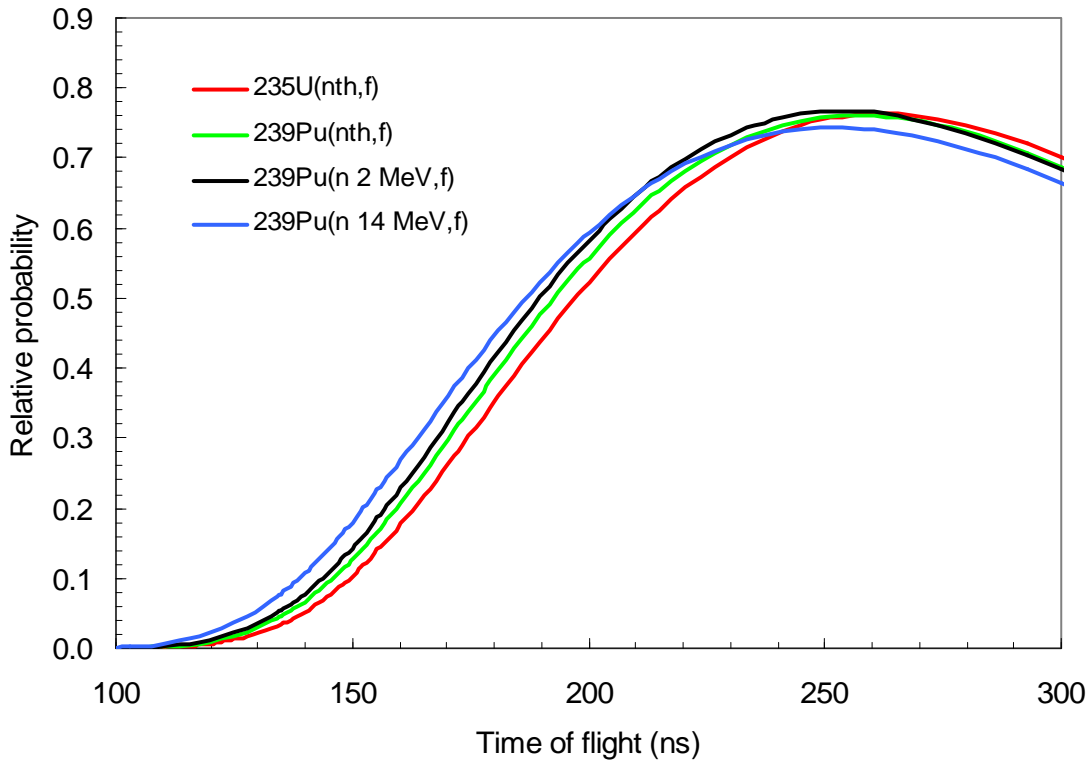

Fig. 3. Same as Fig. 2, but with an expanded time scale to show the differences in the region of the leading edges.

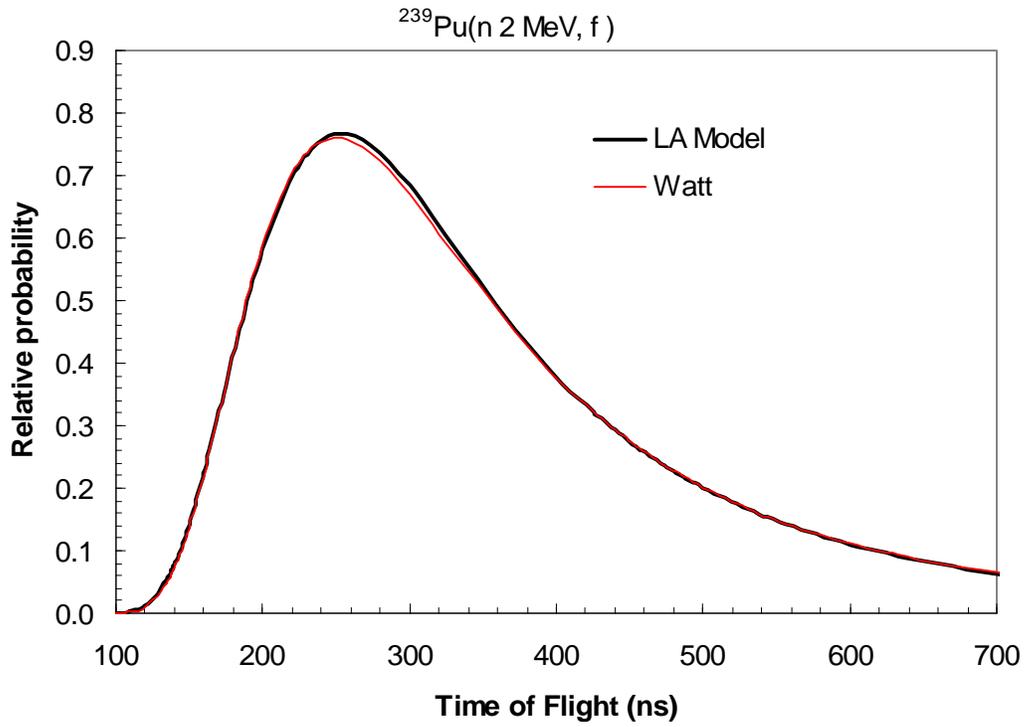

Fig. 4 The Los Alamos Model 2-MeV neutron induced fission of $^{239}$Pu time-of-flight spectrum and the corresponding Watt fit.

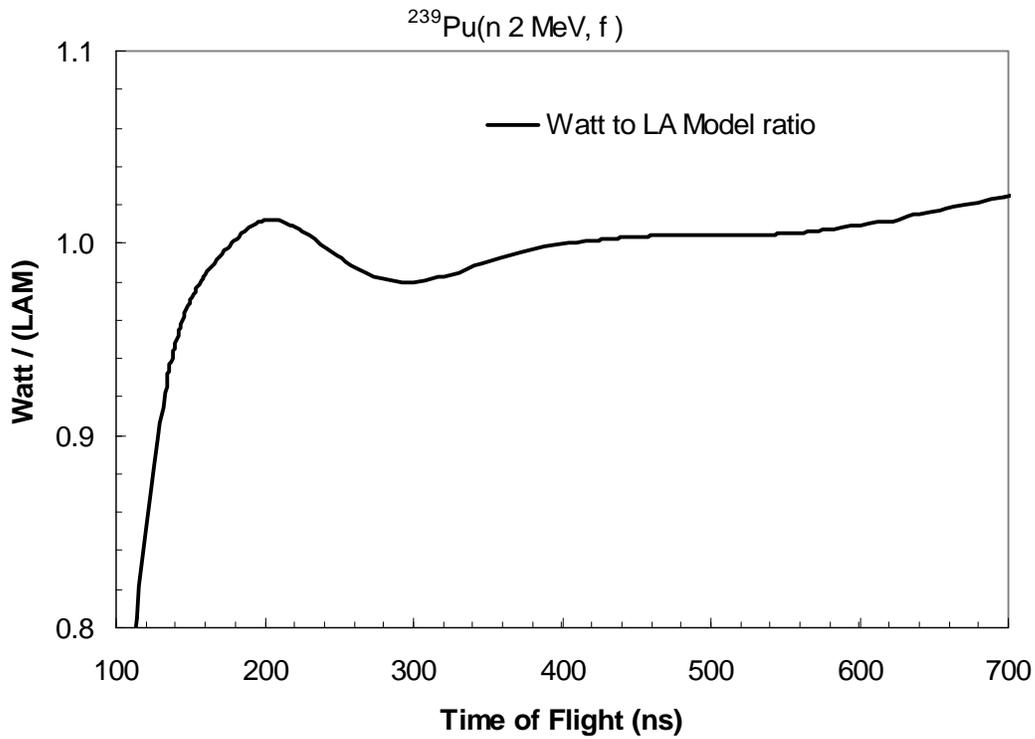

Fig. 5 The ratio of the Watt fit to the Los Alamos Model, for the fit shown in fig. 4.

Table I. Watt fitting parameters

| | $^{235}$U(n,f) | | $^{238}$U(n,f) | | $^{239}$Pu(n,f) | |
|---|---|---|---|---|---|---|
| E (MeV) | $a$ (MeV) | $b$ (MeV$^{-1}$) | $a$ (MeV) | $b$ (MeV$^{-1}$) | $a$ (MeV) | $b$ (MeV$^{-1}$) |
| 0.0 | 0.9235 | 3.0271 | 0.8889 | 2.9408 | 1.0107 | 2.2770 |
| 0.5 | 0.9670 | 2.4862 | 0.9000 | 2.8542 | 1.0210 | 2.2214 |
| 1.0 | 0.9785 | 2.4159 | 0.9111 | 2.7715 | 1.0313 | 2.1679 |
| 1.5 | 0.9755 | 2.4343 | 0.9220 | 2.6926 | 1.0415 | 2.1163 |
| 2.0 | 0.9767 | 2.4267 | 0.9329 | 2.6172 | 1.0515 | 2.0665 |
| 2.5 | 0.9623 | 2.5157 | 0.9436 | 2.5450 | 1.0615 | 2.0189 |
| 3 | 0.9677 | 2.4823 | 0.9543 | 2.4760 | 1.0714 | 1.9729 |
| 4 | 0.9829 | 2.3895 | 0.9753 | 2.3463 | 1.0910 | 1.8856 |
| 5 | 1.0193 | 2.1856 | 0.9960 | 2.2269 | 1.1104 | 1.8042 |
| 6 | 1.0816 | 1.7209 | 1.0878 | 1.2787 | 1.2389 | 0.8098 |
| 7 | 1.2143 | 0.6294 | 1.2543 | 0.01 | 1.4146 | 0.01 |
| 8 | 1.3413 | 0.0100 | 1.2715 | 0.01 | 1.4110 | 0.01 |
| 9 | 1.3308 | 0.0100 | 1.2900 | 0.01 | 1.4194 | 0.01 |
| 10 | 1.3329 | 0.0100 | 1.3056 | 0.01 | 1.4321 | 0.01 |
| 11 | 1.2418 | 0.4576 | 1.3268 | 0.01 | 1.4462 | 0.01 |
| 12 | 1.2117 | 0.7016 | 1.2623 | 0.3258 | 1.4579 | 0.01 |
| 13 | 1.2528 | 0.5045 | 1.2770 | 0.2850 | 1.4624 | 0.01 |
| 14 | 1.3423 | 0.0100 | 1.3301 | 0.01 | 1.4657 | 0.01 |
| 15 | 1.3157 | 0.0100 | 1.3023 | 0.01 | 1.4723 | 0.01 |
| 16 | | | 1.2892 | 0.01 | 1.2234 | 0.8780 |
| 17 | | | 1.2735 | 0.01 | | |
| 18 | | | 1.2700 | 0.01 | | |
| 19 | | | 1.2405 | 0.01 | | |
| 20 | | | 1.1890 | 0.01 | | |

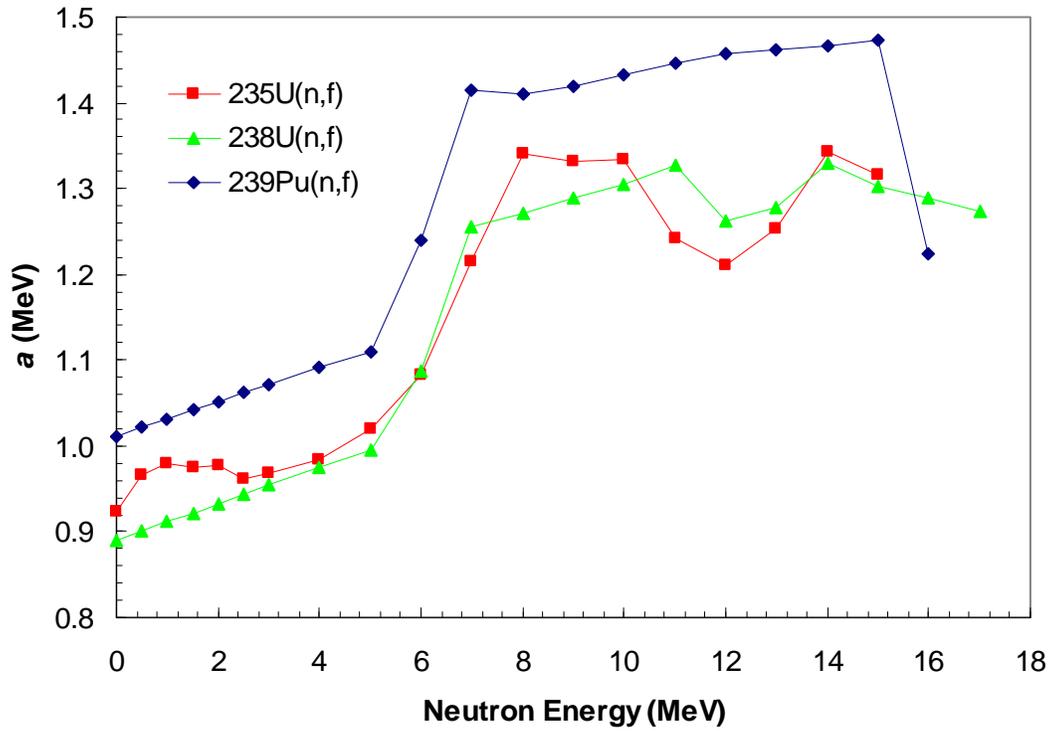

Fig. 6. Watt fitting parameter *a* versus incoming neutron energy.

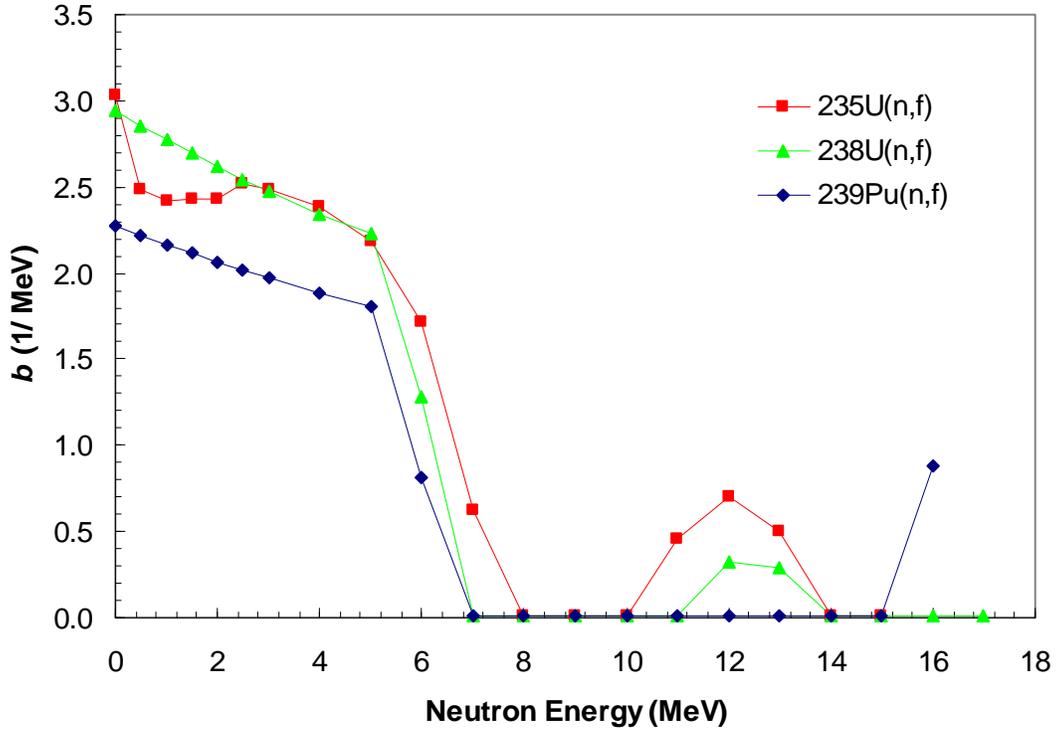

Fig. 7. Watt fitting parameter *b* versus incoming neutron energy.

LA-UR-07-6090

## References

[1] MCNP – A General Monte Carlo N-Particle Transport Code, Version 5, LA-UR-03-1987.
[2] http:/t2.lanl.gov/data/fspect
[3] Madland D. G., and Nix J. R., Nucl. Sci. and Eng. **81**, 213 (1982).

Appendix A : The subroutine getab

```
          subroutine getab(ia,en,a,b,wc)
          implicit none
c         This code was written by John P. Lestone 10/31/06.
c         This subroutine gets Watt parameters for the outgoing fission neutron energy spectrum
c         for 235U(n,f), 238U(n,f), and 239Pu(n,f) as a function of incoming (inducing) neutron kinetic energy.
c         The a's and b's were obtained using the Los Alamos Model fission neutron energy spectrum available at
c         http://t2.lanl.gov/data/fspect/ . Here we use a and b obtained by minimizing a chi-squred in time of flight
c         space. Therefore the a and b obtained using this routine should be used when trying to simulate time of
c         flight data. Different values for a and b might be obtained if the fitting was performed in neutron
c         energy space.
c         The fitting procedure was performed as follows for each fissioning isotope and each inducing neutron
c         energy
c         (1) Each of the fission neutron kinetic energies listed (see http://t2.lanl.gov/data/fspect/ ) were converted
c             into a time of flight (TOF) assuming an ~600 cm flight path.
c         (2) Each corresponding intensity listed at http://t2.lanl.gov/data/fspect/ was scaled by the
c         corresponding neutron energy raised to the power of 3/2. This maps the intensities from energy to TOF
c         space, assuming classical physics.
c         Relativistic corrections are ignored in the energy to TOF mapping. This is justified because the
c         relativistic corrections are small for 14 MeV neutrons (<2%) and the same classical mapping is applied
c          to both the LA Model and the Watt fit. In this case two wrongs do make a right.
c         (3) For each TOF the corresponding Watt values in TOF space were determined for a given choice of "a"
c         and "b", this is given by c x exp(-e/a) x sinh ( sqrt( b x e) ) x e**(3/2),
c         where c = 2 x exp ( -a x b / 4 ) / sqrt ( pi x a**3 x b ).
c         (4) Chi-squared = sum ( LA Model in TOF spaces - Watt spectrum in TOF space )**2.
c                 The sum was from a TOF corresponding to 14 MeV neutrons to 700 ns (~0.37 MeV).
c         (5) "a" and "b" were then adjusted to minimize the chi-squared.
c         This procedure gives Watt parameter values (a and b) at the discrete inducing neutron energies given at
c         http://t2.lanl.gov/data/fspect/. For inducing neutron kinetic energies between the values listed at
c         http://t2.lanl.gov/data/fspect/ we assume linear interpolation of both the "a" and "b" parameters.
c         wc are the Watt coefficients used in MCNP to sample the Watt distribution. These wc(1:3) are simple
c         functions of "a" and "b".
c         To test the wc(1:3), 1E7 neutron energies were sampled using the following piece of code with
c         a=1.051539 and b=2.06653.
c  20     t1=-log(ranf(0.d0))
c         if((log(ranf(0.d0))+wc(1)*(1.+t1))**2.gt.wc(2)*t1) goto 20
c         erg=wc(3)*t1
c         Histogramed results were compared to the corresponding Watt spectrum and found to be in agreement.

c         input parameters
          integer ia   ! ia = 235,238, or 239 for 235U,238U,239Pu(n,f)
          real*8 en    ! energy of the neutron inducing fission (MeV)
c         output parameters
          real*8 a,b   ! Watt parameters, a(MeV), b(1/MeV)
          real*8 wc(1:3) ! Watt coefficients used in MCNP sampling of the Watt distribution
c         internal variables
          real*8 a239 ! function, returns 239Pu(n,f) Watt "a" values as a function of en
          real*8 b239 ! function, returns 239Pu(n,f) Watt "b" values as a function of en
          real*8 a235 ! function, returns 235U (n,f) Watt "a" values as a function of en
          real*8 b235 ! function, returns 235U (n,f) Watt "b" values as a function of en
          real*8 a238 ! function, returns 238U (n,f) Watt "a" values as a function of en
          real*8 b238 ! function, returns 238U (n,f) Watt "b" values as a function of en
c         dummy varables
          real*8 x

          if (ia.eq.235) then
           a=a235(en)
           b=b235(en)
```

```fortran
      elseif (ia.eq.238) then
       a=a238(en)
       b=b238(en)
      else
       a=a239(en)
       b=b239(en)
      endif

      x=1.+0.125*a*b
      wc(1)=x+sqrt(x**2-1.)-1.
      wc(2)=a*b*(wc(1)+1.)
      wc(3)=a*(wc(1)+1.)

      end
c     ************************************************************************************

      function a239(en)
c     This function returns 239Pu(n,f) Watt "a" values as a function of en
      implicit none
      real*8 a239,en
      real*8 e(20),a(20)
      integer i,n
      data e/ 0.0, 0.5, 1.0, 1.5, 2.0, 2.5, 3.0, 4.0, 5.0, 6.0,
     +  7.0, 8.0, 9.0,10.0,11.0,12.0,13.0,14.0,15.0,16.0/
      data a/1.010732,1.021048,1.031282,1.041452,1.051539,
     +  1.061522,1.071430,1.091042,1.110367,1.238907,
     +  1.414576,1.411038,1.419392,1.432086,1.446178,
     +  1.457925,1.462355,1.465668,1.472250,1.223430/

      n=20
      if (en.lt.e(1)) then
       a239=a(1)
      elseif (en.gt.e(n-1)) then
       a239=a(n)
      else
       do i=2,n-1
        if (en.le.e(i)) then
         a239=a(i-1)+(en-e(i-1))*(a(i)-a(i-1))/(e(i)-e(i-1))
         return
        endif
       enddo
      endif

      end
c     ************************************************************************************

      function b239(en)
c     This function returns 239Pu(n,f) Watt "b" values as a function of en
      implicit none
      real*8 b239,en
      real*8 e(20),b(20)
      integer i,n
      data e/ 0.0, 0.5, 1.0, 1.5, 2.0, 2.5, 3.0, 4.0, 5.0, 6.0,
     +  7.0, 8.0, 9.0,10.0,11.0,12.0,13.0,14.0,15.0,16.0/
      data b/2.277001,2.221405,2.167910,2.116257,2.066530,
     +  2.018872,1.972900,1.885588,1.804157,0.809774,
     +  0.010000,0.010000,0.010000,0.010000,0.010000,
     +  0.010000,0.010000,0.010000,0.010000,0.877967/

      n=20
      if (en.lt.e(1)) then
       b239=b(1)
      elseif (en.gt.e(n-1)) then
       b239=b(n)
      else
       do i=2,n-1
        if (en.le.e(i)) then
         b239=b(i-1)+(en-e(i-1))*(b(i)-b(i-1))/(e(i)-e(i-1))
```

```
         return
        endif
       enddo
      endif

      end
c     ********************************************************************************

      function a235(en)
c     This function returns 235U(n,f) Watt "a" values as a function of en
      implicit none
      real*8 a235,en
      real*8 e(19),a(19)
      integer i,n
      data e/ 0.0, 0.5, 1.0, 1.5, 2.0, 2.5, 3.0, 4.0, 5.0, 6.0,
   +   7.0, 8.0, 9.0,10.0,11.0,12.0,13.0,14.0,15.0/
      data a/0.923533,0.967018,0.978505,0.975450,0.976710,
   +   0.962337,0.967650,0.982921,1.019257,1.081624,
   +   1.214333,1.341322,1.330779,1.332881,1.241807,
   +   1.211650,1.252784,1.342330,1.315731/

      n=19
      if (en.lt.e(1)) then
       a235=a(1)
      elseif (en.gt.e(n)) then
       a235=a(n)
      else
       do i=2,n
        if (en.le.e(i)) then
         a235=a(i-1)+(en-e(i-1))*(a(i)-a(i-1))/(e(i)-e(i-1))
         return
        endif
       enddo
      endif

      end
c     ********************************************************************************

      function b235(en)
c     This function returns 235U(n,f) Watt "b" values as a function of en
      implicit none
      real*8 b235,en
      real*8 e(19),b(19)
      integer i,n
      data e/ 0.0, 0.5, 1.0, 1.5, 2.0, 2.5, 3.0, 4.0, 5.0, 6.0,
   +   7.0, 8.0, 9.0,10.0,11.0,12.0,13.0,14.0,15.0/
      data b/3.027099,2.486231,2.415850,2.434316,2.426680,
   +   2.515661,2.482288,2.389466,2.185600,1.720871,
   +   0.629432,0.010000,0.010000,0.010000,0.457588,
   +   0.701582,0.504472,0.010000,0.010000/

      n=19
      if (en.lt.e(1)) then
       b235=b(1)
      elseif (en.gt.e(n-1)) then
       b235=b(n)
      else
       do i=2,n
        if (en.le.e(i)) then
         b235=b(i-1)+(en-e(i-1))*(b(i)-b(i-1))/(e(i)-e(i-1))
         return
        endif
       enddo
      endif

      end
c     ********************************************************************************
```

```fortran
      function a238(en)
c     This function returns 238U(n,f) Watt "a" values as a function of en
      implicit none
      real*8 a238,en
      real*8 e(24),a(24)
      integer i,n
      data e/ 0.0, 0.5, 1.0, 1.5, 2.0, 2.5, 3.0, 4.0, 5.0, 6.0,
     +  7.0, 8.0, 9.0,10.0,11.0,12.0,13.0,14.0,15.0,16.0,
     + 17.0,18.0,19.0,20.0/
      data a/0.888876,0.900032,0.911081,0.922027,0.932873,
     + 0.943622,0.954276,0.975311,0.995997,1.087761,
     + 1.254345,1.271463,1.290047,1.305573,1.326836,
     + 1.262261,1.277025,1.330072,1.302299,1.289161,
     + 1.273504,1.269967,1.240538,1.189006/

      n=24
      if (en.lt.e(1)) then
       a238=a(1)
      elseif (en.gt.e(n)) then
       a238=a(n)
      else
       do i=2,n
        if (en.le.e(i)) then
         a238=a(i-1)+(en-e(i-1))*(a(i)-a(i-1))/(e(i)-e(i-1))
         return
        endif
       enddo
      endif

      end

c     ***********************************************************************************

      function b238(en)
c     This function returns 238U(n,f) Watt "b" values as a function of en
      implicit none
      real*8 b238,en
      real*8 e(24),b(24)
      integer i,n
      data e/ 0.0, 0.5, 1.0, 1.5, 2.0, 2.5, 3.0, 4.0, 5.0, 6.0,
     +  7.0, 8.0, 9.0,10.0,11.0,12.0,13.0,14.0,15.0,16.0,
     + 17.0,18.0,19.0,20.0/
      data b/2.940841,2.854197,2.771544,2.692623,2.617194,
     + 2.545044,2.475974,2.346326,2.226935,1.278703,
     + 0.010000,0.010000,0.010000,0.010000,0.010000,
     + 0.325828,0.284998,0.010000,0.010000,0.010000,
     + 0.010000,0.010000,0.010000,0.010000/

      n=24
      if (en.lt.e(1)) then
       b238=b(1)
      elseif (en.gt.e(n-1)) then
       b238=b(n)
      else
       do i=2,n
        if (en.le.e(i)) then
         b238=b(i-1)+(en-e(i-1))*(b(i)-b(i-1))/(e(i)-e(i-1))
         return
        endif
       enddo
      endif

      end
```